# Fabrication of magnetic clusters and rods using electrostatic co-assembly

M. Yan, L. Chevry and J.-F. Berret*
Matière et Systèmes Complexes, UMR 7057 CNRS Université Denis Diderot Paris-VII, Bâtiment Condorcet, 10 rue Alice Domon et Léonie Duquet, 75205 Paris (France)

E-mail: jean-francois.berret@univ-paris-diderot.fr

**Abstract.** Using a novel protocol for mixing oppositely charged colloids and macromolecules, magnetic clusters and rods are fabricated using 10 nm-iron oxide nanoparticles and polymers. Here, we show that as the dispersions undergo the so-called desalting transition, spherical clusters in the range 100 nm – 1 μm form spontaneously upon dialysis or dilution. With a magnetic field applied during the dialysis, a one-dimensional growth of the aggregates is initiated, resulting in the formation of 1 – 100 μm rods of average diameter 200 nm. In this paper, we demonstrate that the nanostructured rods have inherited the properties of the iron oxide particles, namely to be superparamagnetic. We also discuss the dependence of the magnetic properties as a function of the nanoparticle diameter.

## Introduction

The electrostatic complexation between oppositely charged macromolecules and colloids in aqueous solutions has attracted much attention during the last years because it is an efficient way to control the association at the nanometer scale. Since the pioneering work by Bungenberg de Jong on gelatin and arabic gum [1], the complexation has been investigated on various systems, comprising synthetic [2] and biological [3-5] polymers, multivalent counterions [6], surfactant micelles [7,8,9], organic and inorganic nanoparticles [10,7,11-13,14]. Additional control of the co-assembly process was achieved by using polyelectrolyte-neutral diblock copolymers instead of homopolyelectrolytes [15,16]. With copolymers, the complexation was monitored by the molecular weights of the two blocks and resulted in the spontaneous formation of core-shell colloids. This approach provided excellent results. It was shown to be particularly appropriate for the fabrication of cerium or iron oxide clusters in the 20 – 50 nm range [17,18]. This approach also disclosed a critical issue, that is usually not discussed in the literature. The size and morphology of the co-assemblies were found to depend on the mixing process between the initial particle and polymer dispersions [19]. In order to solve this issue, a novel mixing protocol for bringing oppositely charged species together was elaborated [20]. This protocol was inspired from molecular biology techniques developed for the *in vitro* reconstitutions of chromatin [21]. It consisted first in the screening of the electrostatic interactions by bringing the dispersions to high ionic strength (1 M of inorganic salt), and in a second step in the removal of the salt by dialysis or by dilution. In this paper, we have applied this method for the fabrication of spherical and rodlike clusters of superparamagnetic iron oxide nanoparticles. Here we focus on the effects of the desalting rate on the size of the spherical clusters. We also demonstrate that the nanostructured rods have inherited the superparamagnetic properties of the particles. The findings include four batches of nanoparticles characterized by different sizes and dispersities.

**Experimental Section**

The iron oxide nanoparticles were synthesized by alkaline co-precipitation of iron(II) and iron(III) salts and oxidation of the magnetite ($Fe_3O_4$) into maghemite ($\gamma$-$Fe_2O_3$) particles [22]. At the end of the process, the particles were positively charged, resulting in strong electrostatic repulsions between particles and enhanced colloidal stability. The size distribution was determined from transmission electron microscopy (TEM) measurements and could be represented by a log-normal function, with median diameter $D_0$ and polydispersity s. Fig. 1a displays a TEM image of the 9.3 nm particles showing a rather monodisperse population of nanocrystals and Fig. 1b illustrates the log-normal distribution, with polydispersity s = 0.18. Further characterizations of the dispersions using vibrating sample magnetometry and light scattering were performed and we refer to Refs. [23,24] for a full account of the results. In order to improve their colloidal stability in complex biological solvents [24], the cationic particles were coated by poly(acrylic acid) with molecular weight 2000 g mol$^{-1}$ using the precipitation–redispersion process [13]. The thickness of the $PAA_{2K}$ brush tethered on the particle surfaces was estimated at 3 ± 1 nm by dynamical light scattering. The anionically charged NPs have been co-assembled with a cationic–neutral diblock copolymers, referred to as poly (trimethylammonium ethylacrylate)-*b*-poly (acrylamide), and noted $PTEA_{11K}$-*b*-$PAM_{30K}$ [25]. Here, the indices denote the molecular weight of the different blocks. In aqueous solutions at neutral pH, the chains were found to be disperse and in the state of unimers. Dynamic light scattering revealed an hydrodynamic diameter $D_H$ = 11 nm and size exclusion chromatography a polydispersity index $M_n/M_w$ = 1.6 [9].

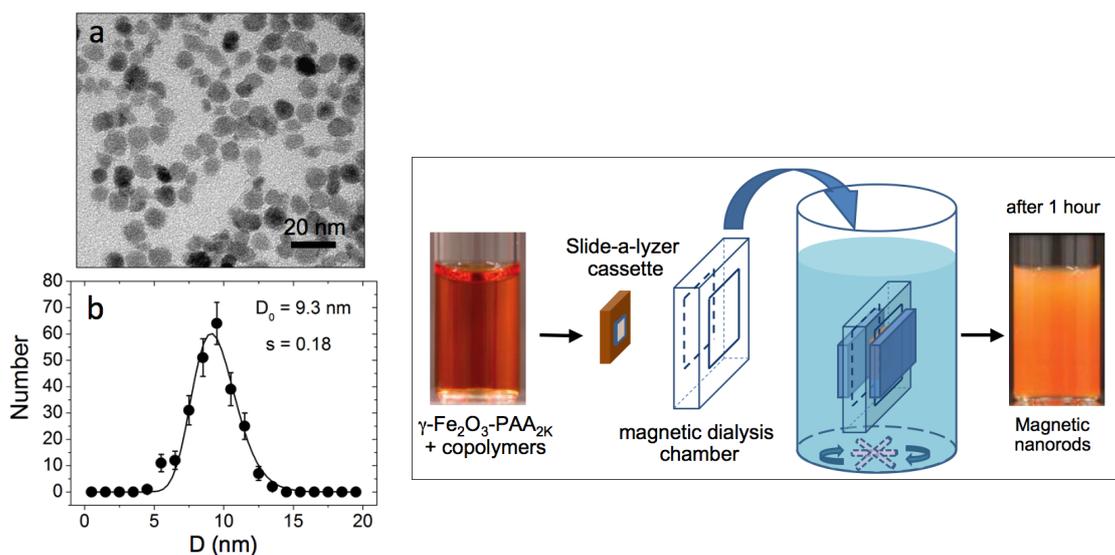

*Figure 1* : *(a) Iron oxide superparamagnetic nanoparticles as observed by transmission electron microscopy (TEM). (b) Size distribution derived from TEM. The continuous line resulted from best fit calculations using a log-normal distribution, with $D_0$ = 9.3 nm and s = 0.18. The diameter is slightly larger than that determined by magnetometry ($D_0$ = 8.3 nm).*
*Figure 2* : *Schematic representation of the protocol that controlled the nanoparticle co-assembly. Without magnetic field, co-assembly yielded spherical clusters, whereas with magnetic field of 0.1 T nanostructured rods in micron range could be fabricated.*

Fig. 2 describes the dialysis protocol that controlled the nanoparticles co-assembly [26]. The strategy involved in a first step the preparation of two separate 1 M $NH_4Cl$ solutions containing respectively the anionic iron oxide particles and the cationic neutral $PTEA_{11K}$-b-$PAM_{30K}$ diblock copolymers. Typical $\gamma$-$Fe_2O_3$ concentrations were 0.1 wt. %, with a polymer-to-nanoparticle volume ratio of 0.5. Both solutions were then mixed with each other and in a second step, the ionic strength of the mixture was progressively diminished by dialysis or dilution. In the dilution process, deionized water was added to mixtures of $PAA_{2K}$-coated nanoparticles and $PTEA_{11K}$-$b$-$PAM_{30K}$ copolymer stepwise. In this case, no magnetic field was applied. In a recent report, we demonstrate that the key parameter that controls the kinetics of formation of electrostatic clusters is the rate $dI_S/dt$ at which the salt is removed from the solution, where $I_S$ denotes the ionic strength. Doing so it was possible to vary $dI_S/dt$ from $10^{-5}$ to 1 M s$^{-1}$. Dialysis was performed against deionized water using a Slide-a-Lyzer® cassette (Thermo Scientific). In these conditions, the whole process reached a stationary and final state within 50 – 100 minutes. The dialysis was carried under two different conditions, with or without magnetic field. The drawing in Fig. 2 represents the case where the magnetic field (0.1 T) was applied. As shown below, the role of the field was to induce the growth of the aggregates in one direction and to favor the formation of the rods. The dialysis experiment between the initial and final ionic strengths was characterized by an average rate of ionic strength change $dI_S/dt \sim 10^{-3}$ - $10^{-4}$ M s$^{-1}$. The aggregates found by dialysis or dilution were further characterized by dynamic light scattering and transmission optical microscopy, using devices described in an earlier report [20].

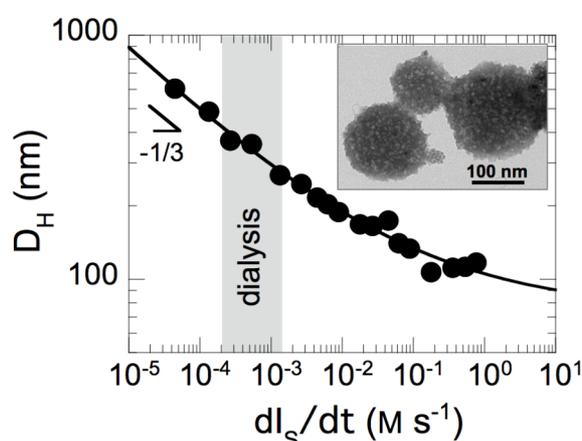

*Figure 3:* Variation of the cluster hydrodynamic diameters $D_H$ as a function of the desalting rate $dI_S/dt$ for slow and rapid dilutions. At low $dI_S/dt$-values, $D_H$ displayed an asymptotic scaling law of the form $D_H \sim (dI_S/dt)^{-1/3}$. Inset : TEM images of spherical aggregates obtained by dialysis. The diameter size of the aggregates was estimated at 180 nm.

## Results and Discussion

*Desalting transition* : In Ref. [19], it was found that with decreasing ionic strength, electrostatically screened dispersions of oppositely charged polymers and nanoparticles underwent an abrupt transition between an unassociated and a clustered state [19]. The transition was dubbed *desalting transition* because it occurred as the excess inorganic salt was removed from the sample. In this former work, the particles put under scrutiny were 7 nm nanoceria, and the polymers were the same $PTEA_{11K}$-$b$-$PAM_{30K}$ as those used here. Moreover, it was shown that the critical ionic strength $I_S^C$ (= 0.42 M) at which the desalting transition took place did not depend on the formulation conditions (such as the concentration or the mixing ratio), nor on the desalting rate $dI_S/dt$. In the present paper, we extended this approach to $PAA_{2K}$-coated iron oxide nanoparticles. In Fig. 3, the desalting transition was monitored on a broad range of desalting rates, lying between $10^{-5}$ and 1 M s$^{-1}$. There, each data point represents the hydrodynamic diameter obtained from a dispersion that underwent the desalting transition. For fast dilution ($dI_S/dt \sim 1$ M s$^{-1}$), the aggregates remained in the 100 nm range. Note that the $D_H$-values in this regime coincide well with those found by direct mixing, suggesting that in terms of kinetics of complexation, the mixing of oppositely charged species is equivalent to a quench [19]. With decreasing $dI_S/dt$, $D_H$ increased and displayed an asymptotic scaling law of the form $D_H \sim dI_S/dt^{-1/3}$. A power law with exponent -1/3 is interesting because it indicates that at infinitely slow dilution, the size of the clusters would diverge, the dispersion exhibiting then a macroscopic phase separation. The inset of Fig. 3 displays a TEM image of nanoparticle aggregates obtained by dialysis. The aggregates of average diameter 180 nm were described as latex-type or composite colloids with a high load of magnetic particles. Assuming a volume fraction of 0.25 inside the large spheres [18], we have estimated that a 200 nm aggregate was built from ~ 6000 particles. Such aggregates were also found to exhibit a remarkable colloidal stability, much probably due to the presence of a neutral poly(acrylamide) brush surrounding the clusters. The data in Fig. 4 are in accordance with those obtained on anionically coated nanoceria [19], showing *in fine* that this protocol can be applied to different nanoparticle systems.

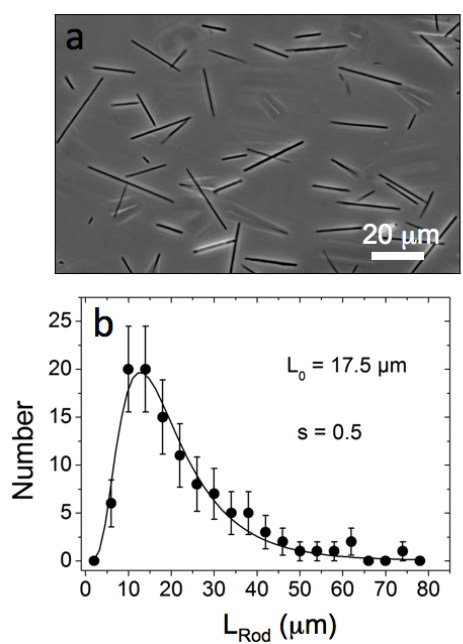

*Figure 4 (right)* : (a) Phase-contrast optical microscopy images (40×) of a dispersion of nanostructured rods made from 8.3 nm γ-$Fe_2O_3$ particles. (b) Experimental and fitted length distribution of the rods displayed in Fig. 4a. The continuous line was obtained from best fit calculations using a log-normal distribution.

*Rod reorientations under 90°-flip of the magnetic field* : In this part, we focus on the results of dialysis experiments performed under the application of a constant magnetic field of 0.1 T (Fig. 2). With a magnetic field applied to the dispersions, a desalting transition identical to that discussed previously for the cluster formation was observed at the critical ionic strength $I_S^c$ = 0.42 M [20]. At the end of the dialysis, once the ionic strength in the cassette reached its equilibrium value around $I_S$ = 5×10$^{-3}$ M, the magnetic field was removed and the dispersion was investigated by optical transmission microscopy. Fig. 4 shows a phase-contrast image obtained with magnification ×40 of randomly oriented and elongated structures with typical sizes in the micrometer range. The γ-$Fe_2O_3$ batch utilized here was made from 8.3 nm particles at concentration c = 0.1 wt. % and polymer-to-nanoparticle volume ratio 0.5. The structures in Fig. 4a were identified as nanostructured rods comprising scores of γ-$Fe_2O_3$ nanoparticles linked together by the cationic polymer "glue". Typical diameters for the rods were found around 200 nm [20]. An image analysis allowed to derive the length distribution of the rods. For this specific sample, it was found to be well accounted for by a log-normal function, with median length $L_0$ = 17.5 ± 0.9 μm and a polydispersity s = 0.50 (Fig. 4b).
In the present work, we investigated the dependence of the magnetic properties of the rods with respect to the diameter of the particles. In a previous work [20], emphasis was put on a unique batch (diameter 7.1 nm) whereas here, particles ranging from 6.7 to 10.7 nm were studied. For that, we performed quantitative measurements of the rod kinetics associated with 90°-reorientations. In such experiments, θ(t) denotes the angle between the major axis of a rod and magnetic excitation H. At the application of the field at 90° with respect to the initial conditions, the nanorods rotated in the plane of observation around their center of gravity. The magnitude of the applied field ranged from 0 to 20 mT, *i.e.* much less than the intensity required to grow the rods (0.1 T).

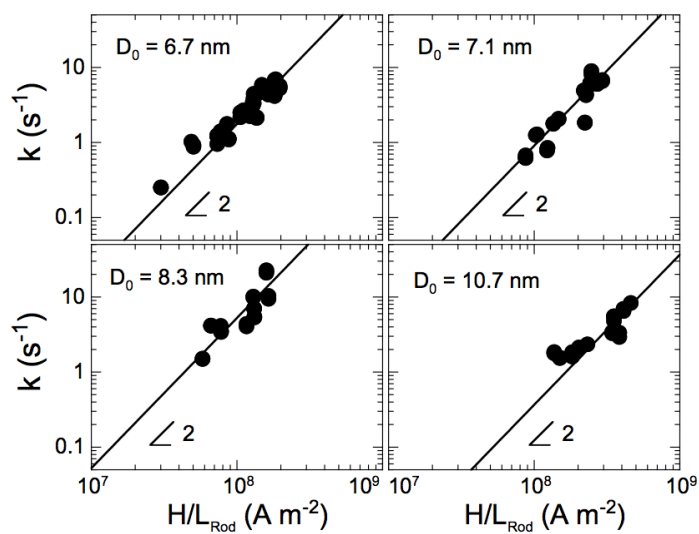

*Figure 5 :* Variation of the parameter k as a function of the ratio $H/L_{Rod}$ for rods made from particle of diameters 6.7 nm, 8.3 nm, 7.1nm and 10.7 nm. k is the slope of the exponential decay, H the external magnetic excitation applied to the rod and $L_{Rod}$ its length. The straight line was computed according to Eq. 2, in agreement with a quadratic dependence $(H/L_{Rod})^2$.

The reorientational dynamics resulted from the balance between the magnetic torque and the hydrodynamic drag, yielding an expression of the form [20] :

$$\tan\theta(t) = \tan\theta_0 \exp(-kt) \quad (1)$$

In Eq. 1, $\theta_0$ was the initial position of the rods prior to the application of the field ($\theta_0 \sim 90°$) and k the decay rate of the reorientation. k depends on the geometrical parameters of the rods, such as the length $L_{Rod}$ and the diameter $D_{Rod}$, and on the strength of the applied field, as [20] :

$$k = \frac{\chi^2}{(2+\chi)} \frac{\mu_0 g(L_{Rod}/D_{Rod})}{2\eta} D_{Rod}^2 \left(\frac{H}{L_{Rod}}\right)^2 \quad (2)$$

Here, $\chi$ denotes the magnetic susceptibility of the rods, $\mu_0$ the permeability of vacuum, $\eta$ the solvent viscosity (0.89×10$^{-3}$ Pa s) and $g(L_{Rod}/D_{Rod})$ a slowly varying function of the aspect ratio. The quadratic H-dependence of the decay rate is a signature of the superparamagnetic character of these aggregates. Figs. 5 display the variations of k as a function of the ratio $H/L_{Rod}$ for rods made from 4 different iron oxide batches. The particle diameters investigated were $D_0$ = 6.7 nm (Fig. 5a), $D_0$ = 7.1 nm (Fig. 5b), $D_0$ = 8.3 nm (Fig. 5c) and $D_0$ = 10.7 nm (Fig. 5d). In the 4 plots, $(H/L_{Rod})^2$-dependences were found in excellent agreement with the predictions of Eq. 2. Note that the data for the 7.1 nm were in quantitative agreement with those reported in Ref. [20]. The value of the prefactor in Eq. 2 allowed us to estimate the magnetic susceptibility $\chi$ as a function of the particle diameter. To this end, the diameter of the rods was assumed to be particle independent, and it was fixed at $D_{Rod}$ = 200 nm for the 4 samples. $\chi$ was found to be respectively 2.5 ± 0.5, 1.6 ± 0.3, 5.6 ± 0.8 and 0.9 ± 0.2. Recent calculations of the susceptibility on concentrated dispersions taking into account the magnetic inter-

Published in: **Progress in Colloid and Polymer Science** (2010) 137:35–39
DOI: 10.1007/2882_2010_8particle interactions have shown that $\chi$ should increase typically as $\chi(D) \sim D_0^3$ [27]. The change in the particle diameters from 6.7 to 10.7 nm should have resulted in a fourfold increase of the susceptibility. This result was not observed in the present work. This discrepancy could stem from the assumption made for the rod diameter. The diameter appears as a quadratic dependence in Eq. 2, and small variations on $D_{Rod}$ from one sample to the other may impact considerably the estimations for $\chi$. These findings finally suggest that the 90°-reorientation kinetic measurements should be performed in parallel with an accurate determination of the rod diameters by TEM.

## Conclusion

In this communication, we have shown that the electrostatic interactions between oppositely charged magnetic nanoparticles and polymers can be controlled by tuning the ionic strength of the dispersion. With decreasing $I_S$, the electrostatically screened polymers and nanoparticles undergo an abrupt transition between an unassociated and a clustered state. By tuning the desalting kinetics $dI_S/dt$ from 1 to $10^{-5}$ M s$^{-1}$, the size of the magnetic clusters was varied from 100 nm to $\sim$ 1 µm. Under the application of a magnetic field, nanostructured rods with lengths in micrometer range were fabricated. We demonstrate here that the rods have inherited the properties of the nanoparticles, namely to be superparamagnetic. This was shown for particles with sizes ranging from 6.7 to 10.7 nm by investigating the 90°-reorientation kinetics of rods. It was found that the decay rate varied as $(H/L_{Rod})^2$, in good agreement with theoretical predictions. The present approach of magnetic nanoparticle co-assembly should open new perspectives for the fabrication of nanodevices such as tips, tweezers and actuators applicable in biophysics and biomedicine.

## Acknowledgement
We thank Jérôme Fresnais, Olivier Sandre and Régine Perzynski for fruitful discussions. The Laboratoire Physico-chimie des Electrolytes, Colloïdes et Sciences Analytiques (PECSA, Université Pierre et Marie Curie, Paris, France) is acknowledged for providing us with the nanoparticles and for the access to the TEM facilities. This research was supported by the Agence Nationale de la Recherche under the contract BLAN07-3_206866, by the European Community through the project : "NANO3T—Biofunctionalized Metal and Magnetic Nanoparticles for Targeted Tumor Therapy", project number 214137 (FP7-NMP-2007-SMALL-1) and by the Région Ile-de-France (DIM project on Health, Environnement and Toxicology).## References

1. Bungenberg-de-Jong MG (1949) in: H.R. Kruyt (Ed.) Elsevier New York 335
2. Laugel N, Betscha C, Winterhalter M, Voegel J-C, Schaaf P, Ball V (2006) J. Phys. Chem. B 110:19443
3. Raspaud E, Olvera-de-la-Cruz M, Sikorav J-L, Livolant F (1998) Biophys. J. 74:381
4. Jeong JH, Kim SW, Park TG (2003) Bioconjugate Chem. 14:473
5. Weinbreck F, de Vries R, Schrooyen P, de Kruif CG (2003) Biomacromolecules 4:293
6. Olvera-de-la-Cruz M, Belloni L, Delsanti M, Dalbiez JP, Spalla O, Drifford M (1995) J. Chem. Phys 103:5781
7. Bronich TK, Cherry T, Vinogradov S, Eisenberg A, Kabanov VA, Kabanov AV (1998) Langmuir 14:6101
8. Zhou S, Yeh F, Burger C, Chu B (1999) J. Phys. Chem. B 103:2107
9. Berret J-F (2005) J. Chem. Phys. 123:164703
10. Biggs S, Scales PJ, Leong Y-K, Healy TW (1995) J. Chem. Soc. Faraday Trans. 91:2921
11. Schneider G, Decher G (2004) Nano Lett. 4:1833
12. Si S, Kotal A, Mandal TK, Giri S, Nakamura H, Kohara T (2004) Chem. Mater. 16:3489
13. Sehgal A, Lalatonne Y, Berret J-F, Morvan M (2005) Langmuir 21:9359